\documentclass[twocolumn]{autart}  

\usepackage{graphicx}

\usepackage{mathtools}

\begin{document}

\begin{frontmatter}

\title{Optimal Input Placement in Lattice Graphs\thanksref{footnoteinfo}}

\thanks[footnoteinfo]{Corresponding author I.~S.~Klickstein.}

\author[UNM]{Isaac Klickstein}\ead{iklick@unm.edu},
\author[UNM]{Francesco Sorrentino}\ead{fsorrent@unm.edu}

\address[UNM]{Department of Mechanical Engineering. 1 University of New Mexico. Albuquerque, NM 87131}

\begin{keyword}
Networks; Linear optimal control; Lattices; Lyapunov equation; Discrete Fourier transforms.
\end{keyword}
%
\begin{abstract}
The control of dynamical, networked systems continues to receive much attention across the engineering and scientific research fields.
  Of particular interest is the proper way to determine which nodes of the network should receive external control inputs in order to effectively and efficiently control portions of the network.
  Published methods to accomplish this task either find a minimal set of driver nodes to guarantee controllability or a larger set of driver nodes which optimizes some control metric.
  Here, we investigate the control of lattice systems which provides analytical insight into the relationship between network structure and controllability.
  First we derive a closed form expression for the individual elements of the controllability Gramian of infinite lattice systems.
  Second, we focus on nearest neighbor lattices for which the distance between nodes appears in the expression for the controllability Gramian.
  We show that common control energy metrics scale exponentially with respect to the maximum distance between a driver node and a target node.
\end{abstract}
\end{frontmatter}
\section{Introduction}
  The control of dynamical networks continues to receive much attention throughout the engineering literature \cite{liu2016control,gao2014target,zhou2015controllability,klickstein2017energy}.
  Typically, the problems explored consist of either finding a minimal set of driver nodes (structural controllability \cite{lin1974structural,liu2011controllability} , exact controllability \cite{yuan2013exact}, among others \cite{cowan2012nodal,pequito2017robust,zhang2017efficient}) or determining the particular control inputs (minimum energy \cite{yan2012controlling,yan2015spectrum}, LQR \cite{lin2011augmented,klickstein2017energy}, proportional \cite{yu2009pinning,tang2014synchronization}, and others).
  The minimum energy control strategy is often investigated as it lower bounds the $\mathcal{L}2$ norm of any other control strategy that performs the same control action (initial condition to final condition).
  The minimum energy control strategy can be characterized by the controllability Gramian, a symmetric positive semi-definite matrix that is a function of the weighted adjacency matrix of the underlying network 
  and the distribution of inputs into the system.\\
  \indent
  It has been shown that the minimal set of driver nodes which ensures controllability, may not be numerically feasible \cite{sun2013controllability}, that is, the resulting control input requires the inversion of a very ill-conditioned matrix.
  To compensate, additional driver nodes may be added to the minimal set either randomly \cite{yan2012controlling,yan2015spectrum,klickstein2017energy} or according to some heuristic \cite{chen2016energy}.
  Alternatively, it has been shown that optimizing the placement of driver nodes with respect to a variety of submodular control energy metrics is NP-hard \cite{olshevsky2014minimal,tzoumas2016minimal} which inspired the development of greedy approximation algorithms \cite{summers2016submodularity,summers2016actuator,tzoumas2015minimal,tzoumas2016minimal}.
  These algorithms require many high accuracy computations of controllability Gramians along with their inverse or determinant to make the necessary decisions in the greedy algorithm which 
  is computationally expensive even when the graph is sparse \cite{benner2013numerical}.\\
  Our first contribution is a closed form expression for each entry of the controllability Gramian for arbitrary infinite lattice networks.
  This framework allows one to investigate the role that various connectivity patterns have on the control energy.\\
  Our second contribution finds the exponential decay of an entry in the Gramian with respect to the distance between the target nodes and the driver nodes in a nearest neighbor lattice graph.
  Using this decay rate, and the Cauchy interlacing theorem, these control metrics in the nearest neighbor lattice graph are lower bounded by a function of the distance between a driver node and the target node furthest away.
  This exponential scaling we derive has previously been observed numerically \cite{chen2016energy,wang2017physical} for the finite chain graph where the authors demonstrated the exponential scaling holds for graphs of a general topology.
  Taken together, driver node placement algorithms that minimize the distance between driver nodes and target nodes by using a discrete location formulation \cite{mirchandani1990discrete} may be a computationally cheaper alternative to the currently available greedy approximation algorithms to choose a driver node set.\\
  \indent
  The remainder of the paper is as follows.
  In section 2 we present the derivation of both the time-varying and the steady state controllability Gramian of a general lattice network.
  In section 3 we specialize the results in section 2 to nearest neighbor lattices.
  In section 4 we discuss the implications of our methodology to developing driver node placement algorithms.
%
%
\section{Preliminaries}
We first define a lattice graph in general, then write the linear dynamics that governs the states of each node.
With the given definitions, we then derive exact formulas for the elements of the controllability Gramian, which contains information about the minimum control energy, or effort, required to perform any required control action.
%
%
\subsection{Lattice Graphs}
Here we define lattice graphs and introduce some notational simplifications which we use in the following derivations.
Note that while typically in the graph theory literature nodes of a graph are labeled with a single positive integer, here we define node labels as  vectors of integers.
%
%
%
\begin{defn}\label{def:lattice}{(Lattice Graphs)}
  A lattice graph $\mathcal{L} = (\mathcal{V},\mathcal{E})$ consists of an infinite set of nodes $\mathcal{V} = \{v_{\textbf{i}}|\textbf{i} \in \mathcal{Z}^d\}$ where $\textbf{i}$ is node $v_{\textbf{i}}$'s index and a set of edges $\mathcal{E} \subset \mathcal{V} \times \mathcal{V}$.
  The lattice graph can be completely described by three properties.
  \begin{enumerate}
    \item The dimension of the lattice $d$ is the dimension of the indices of the nodes $\textbf{i} \in \mathcal{Z}^d$, called lattice sites.
    There is a node $v_{\textbf{i}}$ at every lattice site.
    \item The coupling of the lattice is described by a set $\mathcal{N} \subset \mathcal{Z}^d$ such that,
    \begin{equation}
      (v_{\textbf{i}+\textbf{n}},v_{\textbf{i}}) \in \mathcal{E},\ \forall \textbf{n} \in \mathcal{N},\ \forall \textbf{i} \in \mathcal{Z}^d
    \end{equation}
    and $(v_{\textbf{i}+\textbf{n}},v_{\textbf{i}}) \notin \mathcal{E}$ for $\textbf{n} \notin \mathcal{N}$, $\forall \textbf{i} \in \mathcal{Z}^d$.
    The number of incoming edges of every node $v_{\textbf{i}} \in \mathcal{V}$ is $|\mathcal{N}|$.
    Note that a lattice in general is directed.
    \item There is a function $\psi:\mathcal{N} \mapsto \mathcal{R}$ that describes the edge weights.
    This function implies that the weight of edge $(v_{\textbf{i}+\textbf{n}},v_{\textbf{i}})$ is equal to the weight of edge $(v_{\textbf{j}+\textbf{n}},v_{\textbf{j}})$ for all $\textbf{i},\textbf{j} \in \mathcal{Z}^d$, i.e., it is independent of the node index.
    The weights may be positive or negative.
  \end{enumerate}
\end{defn}
%
%
%
\begin{figure}
  \centering
  \includegraphics[width=\columnwidth]{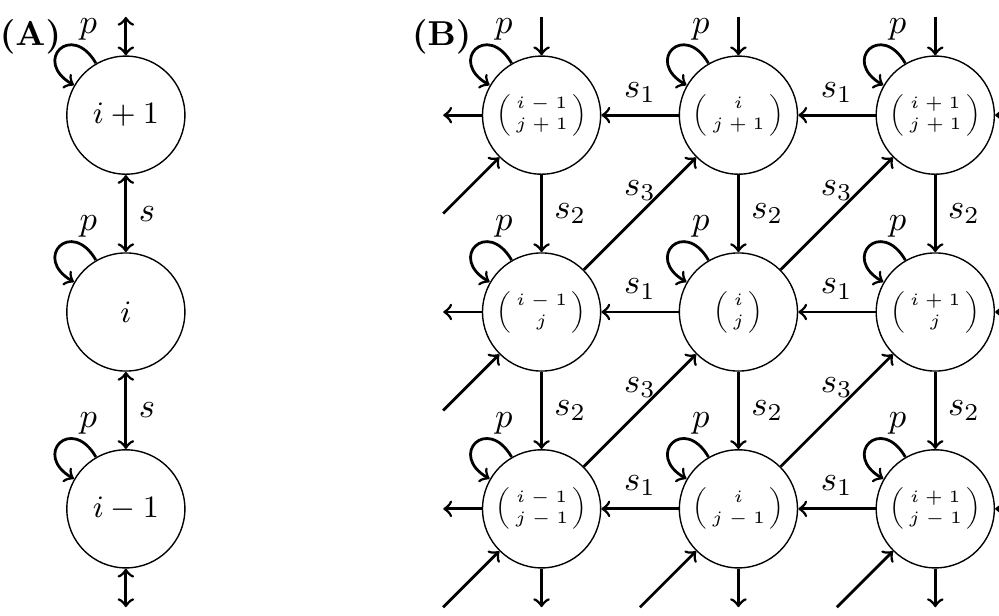}
  \caption{
    Two examples of lattices.
    (A) A one-dimensional lattice with properties $d = 1$, $\mathcal{N} = \{0,-1,1\}$ and edge weights $\psi(0) = p$ and $\psi(-1) = \psi(1) = s$.
    This is the generic form of the one-dimensional nearest neighbor lattice we cover in more detail in section 3.
    (B) A two-dimensional lattice with properties $d = 2$, $\mathcal{N} = \left\{ (0,0),(0,1),(1,0),(-1,-1) \right\}$, and edge weights $\psi(0,0) = p$, $\psi(0,1) = s_1$, $\psi(1,0) = s_2$, $\psi(-1,-1) = s_3$.
  }
  \label{fig:lattice}
\end{figure}
Two examples of lattices are shown in Fig. \ref{fig:lattice}.
The first example in Fig. \ref{fig:lattice}(A) is a one-dimensional nearest neighbor lattice.
Here, each node $v_{\textbf{i}}$ has $|\mathcal{N}|=3$ neighbors; $v_{\textbf{i}-1}$, $v_{\textbf{i}+1}$, and the node itself, $v_{\textbf{i}}$ for every $\textbf{i} \in \mathcal{Z}$.
The second example in Fig. \ref{fig:lattice}(B) is a two-dimensional lattice with directed edges listed in the caption.
We emphasize that the remaining results in this section are applicable to any lattice of arbitrary dimension and connectivity pattern that can be described by Definition \ref{def:lattice}.\\
\indent
We also find it useful to define some additional operations on integer vectors in order to maintain clarity in some of the following equations.
%
%
\begin{defn}{(Operations on Integer Tuples)}
  \begin{itemize}
    \item Let the product of two indices be defined elementwise, that is, the $k$th element of the product $(\textbf{i} \textbf{j})_k = i_k j_k$.
    \item Let the Dirac delta function between two indices be defined as $\delta_{\textbf{i},\textbf{j}} = \prod_{k=1}^d \delta_{i_k,j_k}$ where $\delta_{i_k,j_k} = 1$ if $i_k = j_k$ and $\delta_{i_k,j_k} = 0$ otherwise.
    \item Let the exponential of an index be $e^{a \textbf{i}} = \prod_{k=1}^d e^{a i_k}$ where $a$ is a complex coefficient.
    \item Let the integral with respect to an integer vector be defined as,
    \begin{equation}
      \int_a^b f(\textbf{i}) d\textbf{i} = \int_a^b \cdots \int_a^b f(\textbf{i}) d i_1 \ldots d i_d
    \end{equation}
  \end{itemize}
\end{defn}
%
%
\subsection{Linear Dynamics}
Each node $v_{\textbf{i}}$ in our lattice $\mathcal{L}$ is endowed with a time-varying state $x_{\textbf{i}}(t) \in \mathcal{R}$.
These time-varying states may represent position or velocity for formation type problems \cite{yu2010some}, or transfer rates in routing problems \cite{zhao2005onset}, or average excitement in neuronal networks \cite{wang2015passivity}, and many other network applications.
Before presenting the dynamical system that describes the time evolution of the states of each node, we will introduce two finite subsets of the nodes in the lattice.
%
%
%
\begin{defn}[Driver Nodes]
  Let $\mathcal{D} \subset \mathcal{V}$ be the set of driver nodes of cardinality $|\mathcal{D}| = n_d$.
  If node $v_{\textbf{i}} \in \mathcal{D}$, then node $v_{\textbf{i}}$ receives an external control input $u_{\textbf{i}}(t)$ which we are free to define.
  We assume that no control input can be connected to more than a single node.
\end{defn}
%
%
\begin{defn}{(Target Nodes)}
  Let $\mathcal{T} \subset \mathcal{V}$ be the set of target nodes of cardinality $|\mathcal{T}| = n_t$.
  If $v_{\textbf{i}} \in \mathcal{T}$, then there is a desired value for the state $x_{\textbf{i}}(t_f) = x_{\textbf{i},f}$ which is chosen before applying the control action.
\end{defn}
\begin{rem}
  Note that we will assume both $\mathcal{D}$ and $\mathcal{T}$ are finite subsets of $\mathcal{V}$ and they may overlap, that is, it is possible for a node $v_{\textbf{i}} \in \mathcal{D} \cup \mathcal{T}$.
\end{rem}
%

%
The linear differential equation that governs the time evolution of the each state is,
\begin{equation}\label{eq:sys}
  \dot{x}_{\textbf{i}}(t) = \sum_{\textbf{n} \in \mathcal{N}} \psi(\textbf{n}) x_{\textbf{i}+\textbf{n}}(t) + \sum_{v_{\textbf{a}} \in \mathcal{D}} \delta_{\textbf{i},\textbf{a}} u_{\textbf{a}}(t).
\end{equation}
While here we consider scalar dynamics ($x_\textbf{i}(t) \in \mathcal{R}$ and $\psi(\textbf{n}) \in \mathcal{R}$), the extension to multi-dimensional states, $\textbf{x}_{\textbf{i}}(t) \in \mathcal{R}^{n_x}$ and $\boldsymbol{\psi}(\textbf{n}) \in \mathcal{R}^{n_x \times n_x}$ is straightforward.
Note that the time evolution of the states of each node $v_{\textbf{i}}$ are a function of its immediate neighbors $v_{\textbf{i}+\textbf{n}}$ for all $\textbf{n} \in \mathcal{N}$ and possibly a driver node if $v_{\textbf{i}} \in \mathcal{D}$.
Each node has an initial condition $x_{\textbf{i}}(0) = x_{\textbf{i},0}$ for all $v_{\textbf{i}} \in \mathcal{V}$, and the target nodes have a final condition $x_{\textbf{i}}(t_f) = x_{\textbf{i},f}$ for all $v_{\textbf{i}} \in \mathcal{T}$.
We would like to achieve this final condition with minimal control energy.
\begin{equation}\label{eq:opt}
  \begin{aligned}
    \min && &J = \frac{1}{2} \int_0^{t_f} \sum_{v_{\textbf{a}} \in \mathcal{D}} u_{\textbf{a}}^2(\tau) d\tau\\
    \text{s.t.} && &\dot{x}_{\textbf{i}}(t) = \sum_{\textbf{n} \in \mathcal{N}} \psi(\textbf{n}) x_{\textbf{i}+\textbf{n}}(t) + \sum_{v_{\textbf{a}} \in \mathcal{D}} \delta_{\textbf{i},\textbf{a}} u_{\textbf{a}}(t), \quad v_{\textbf{i}} \in \mathcal{V}\\
    && &x_{\textbf{i}}(0) = x_{\textbf{i},0}, \quad v_{\textbf{i}} \in \mathcal{V}, \quad x_{\textbf{i}}(t_f) = x_{\textbf{i},f}, \quad v_{\textbf{i}} \in \mathcal{T}
  \end{aligned}
\end{equation}
The optimal cost of this optimal control problem can be written \cite{klickstein2017energy},
\begin{equation}\label{eq:optcost}
  J^* = \frac{1}{2} \textbf{b}^T \bar{W}^{-1} \textbf{b}
\end{equation}
The vector $\textbf{b}$ has entries representing the difference between the prescribed final condition $x_{\textbf{i},f}$ and what the state would be without any external control input.
The matrix $\bar{W}$ is the output controllability Gramian \cite{klickstein2017energy} which is a principal submatrix of the controllability Gramian corresponding to the indices of the target nodes.
The elements of the controllability Gramian obey the Lyapunov equation, 
\begin{equation}\label{eq:Wdot}
  \begin{aligned}
    \dot{W}_{\textbf{i},\textbf{j}}(t) &= \sum_{\textbf{n} \in \mathcal{N}} \psi(\textbf{n}) W_{\textbf{i}+\textbf{n},\textbf{j}}(t) + \sum_{\textbf{n}\in\mathcal{N}} \psi(\textbf{n}) W_{\textbf{i},\textbf{j}+\textbf{n}}(t)\\
    &+ \sum_{v_{\textbf{a}} \in \mathcal{D}} \delta_{\textbf{i},\textbf{a}} \delta_{\textbf{j},\textbf{a}}.
  \end{aligned}
\end{equation}
The solution of Eq. \eqref{eq:Wdot} for systems described by lattice graphs is derived by applying the $2d$-dimension discrete time Fourier transform.
\begin{thm}
  Let $\mathcal{I} = \sqrt{-1}$, the unit complex value, and let,
  \begin{equation}
    \phi(\textbf{k}) = \sum_{\textbf{n} \in \mathcal{N}} \psi(\textbf{n}) e^{-\mathcal{I} \textbf{n} \textbf{k}}
  \end{equation}
  where we call $\phi(\cdot) : \mathcal{Z}^d \mapsto \mathcal{C}$ the lattice function for lattice $\mathcal{L}$.
  The time evolution of the controllability Gramian entries for the infinite lattice is,
  \begin{equation}\label{eq:Wt}
    \begin{aligned}
      W_{\textbf{i},\textbf{j}}(t) &= \frac{1}{(2\pi)^{2d}} \sum_{v_{\textbf{a}} \in \mathcal{D}} \int_{-\pi}^{\pi} \int_{-\pi}^{\pi} e^{-\mathcal{I}(\textbf{i}-\textbf{a}) \hat{\textbf{i}}} e^{-\mathcal{I} (\textbf{j}-\textbf{a}) \hat{\textbf{j}}}\\
      &\times \theta(t,\hat{\textbf{i}},\hat{\textbf{j}}) d\hat{\textbf{i}} d\hat{\textbf{j}} 
    \end{aligned}
  \end{equation}
  where the time varying portion,
  \begin{equation}
    \begin{aligned}
    \theta(t,\hat{\textbf{i}},\hat{\textbf{j}})
    = \frac{\exp{\left[\left(\phi(\hat{\textbf{i}}) + \phi(\hat{\textbf{j}})\right)t\right]} - 1}{\phi(\hat{\textbf{i}}) + \phi(\hat{\textbf{j}})}
    \end{aligned}
  \end{equation}
\end{thm}
\begin{pf*}{Proof.}
  First, we state two facts:
  \begin{itemize}
    \item Let $a$ be a complex number. Then $a \int_0^t e^{a\tau} d\tau = e^{at} - 1$.
    \item Let $m$ be an integer.
    An important integral that appears is $\int_{-\pi}^{\pi} e^{-\mathcal{I} m x} dx = 2 \pi \delta_{m,0}$
  \end{itemize}
  From Eq. \eqref{eq:Wt}, it is simple to show that,
  \begin{equation}\label{eq:sumW}
    \begin{aligned}
      \sum_{\textbf{n}\in\mathcal{N}} \psi(\textbf{n}) W_{\textbf{i}+\textbf{n},\textbf{j}}
      &= \frac{1}{(2\pi)^{2d}} \sum_{v_{\textbf{a}} \in \mathcal{D}} \int_{-\pi}^{\pi} \int_{-\pi}^{\pi} \phi(\hat{\textbf{i}})\\
      &\times e^{-\mathcal{I} (\textbf{i}-\textbf{a})\hat{\textbf{i}}} e^{-\mathcal{I}(\textbf{j}-\textbf{a}) \hat{\textbf{j}}} \theta(t,\hat{\textbf{i}},\hat{\textbf{j}}) d\hat{\textbf{i}}d\hat{\textbf{j}}
    \end{aligned}
  \end{equation}
  Applying Eq. \eqref{eq:sumW} and the two facts stated above to Eq. \eqref{eq:Wdot}, we can complete the proof,
  \begin{equation}
    \begin{aligned}
      &\dot{W}_{\textbf{i},\textbf{j}}(t) = 
      %
      %
       \frac{1}{(2\pi)^{2d}} \sum_{v_{\textbf{a}} \in \mathcal{D}} \int_{-\pi}^{\pi} \int_{-\pi}^{\pi} e^{-\mathcal{I}(\textbf{i}-\textbf{a})\hat{\textbf{i}}} e^{-\mathcal{I} (\textbf{j}-\textbf{a})\hat{\textbf{j}}}\\
      &\times\left( e^{(\phi(\hat{\textbf{i}}) + \phi(\hat{\textbf{j}}))t} - 1 \right) d\hat{\textbf{i}} d\hat{\textbf{j}} + \sum_{v_{\textbf{a}} \in \mathcal{D}} \delta_{\textbf{i},\textbf{a}} \delta_{\textbf{j},\textbf{a}}\\
      &= \dot{W}_{\textbf{i},\textbf{j}}(t) - \frac{1}{(2\pi)^{2d}} \sum_{v_{\textbf{a}} \in \mathcal{D}} \int_{-\pi}^{\pi} \int_{-\pi}^{\pi}e^{-\mathcal{I}(\textbf{i}-\textbf{a}) \hat{\textbf{i}}}\\
      &\times e^{-\mathcal{I}(\textbf{j}-\textbf{a})\hat{\textbf{j}}} d\hat{\textbf{i}} d\hat{\textbf{j}} + \sum_{v_{\textbf{a}} \in \mathcal{D}} \delta_{\textbf{i},\textbf{a}} \delta_{\textbf{j},\textbf{a}}\\
      &= \dot{W}_{\textbf{i},\textbf{j}}(t) - \sum_{v_{\textbf{a}} \in \mathcal{D}} \prod_{k=1}^d \left[ \frac{1}{2\pi} \int_{-\pi}^{\pi} e^{-\mathcal{I}(i_k-a_k)\hat{i}_k} d\hat{i}_k \right]\\
      &\times\left[ \frac{1}{2\pi} \int_{-\pi}^{\pi} e^{-\mathcal{I}(j_k-a_k)\hat{j}_k} d\hat{j}_k \right] + \sum_{v_{\textbf{a}} \in \mathcal{D}} \delta_{\textbf{i},\textbf{a}} \delta_{\textbf{j},\textbf{a}}\\
      &= \dot{W}_{\textbf{i},\textbf{j}}(t) - \sum_{v_{\textbf{a}} \in \mathcal{D}} \delta_{\textbf{i},\textbf{a}} \delta_{\textbf{j},\textbf{a}} + \sum_{v_{\textbf{a}} \in \mathcal{D}} \delta_{\textbf{i},\textbf{a}} \delta_{\textbf{j},\textbf{a}}
      = \dot{W}_{\textbf{i},\textbf{j}}(t)
    \end{aligned}
  \end{equation}
\end{pf*}
\begin{rem}
  Of particular interest to us is the case when $\phi(\textbf{k}) < 0$ because then there exists a steady state solution to Eq. \eqref{eq:Wdot},
  \begin{equation}\label{eq:W}
    \begin{aligned}
    &W_{\textbf{i},\textbf{j}} = \lim_{t \rightarrow \infty} W_{\textbf{i},\textbf{j}}(t)\\
    &= -\frac{1}{(2\pi)^{2d}} \sum_{v_{\textbf{a}} \in \mathcal{D}} \int_{-\pi}^{\pi} \int_{-\pi}^{\pi} \frac{e^{-\mathcal{I}(\textbf{i}-\textbf{a})\hat{\textbf{i}}} e^{-\mathcal{I}(\textbf{j}-\textbf{a})\hat{\textbf{j}}}}{\phi(\hat{\textbf{i}}) + \phi(\hat{\textbf{j}})} d\hat{\textbf{i}} d\hat{\textbf{j}}.
    \end{aligned}
  \end{equation}
  When the dynamical system is stable, the minimum energy expression in Eq. \eqref{eq:optcost} approaches a constant value.
  Note that $\phi(\textbf{k}) < 0$ implies $\boldsymbol{0} \in \mathcal{N}$ and $\psi(\boldsymbol{0}) < \sum_{\textbf{n} \in \mathcal{N},\textbf{n} \neq \boldsymbol{0}} \psi(\textbf{n})$, i.e., there exists a large enough, and negative, self-loop at each node.
\end{rem}
From Eq. \eqref{eq:Wdot}, we know $W_{\textbf{i},\textbf{j}}(t)$ is a real number and so the complex portion of both Eq. \eqref{eq:Wt} and \eqref{eq:W} must be zero.
With this in mind, Eq. \eqref{eq:Wt} can be rewritten as,
\begin{equation}\label{eq:Wt_real}
  \begin{aligned}
    W_{\textbf{i},\textbf{j}}(t) &= \frac{1}{(2\pi)^{2d}} \sum_{v_{\textbf{a}} \in \mathcal{D}} \int_{-\pi}^{\pi} \int_{-\pi}^{\pi} \left[ \frac{(\alpha_{\textbf{a}} \sigma + \beta_{\textbf{a}} \omega)(r(t)-1)}{\sigma^2 + \omega^2} \right.\\
    &\left.- \frac{(\beta_{\textbf{a}}\sigma - \alpha_{\textbf{a}}\omega) s(t)}{\sigma^2 + \omega^2} \right] d\hat{\textbf{i}} d\hat{\textbf{j}}
  \end{aligned}
\end{equation}
where the functions are,
\begin{equation}
  \begin{aligned}
    r(t) &= e^{\sigma t} \cos \omega t, \quad
    s(t) = e^{\sigma t} \sin \omega t\\
    \sigma &= \sum_{\textbf{n} \in \mathcal{N}} \psi(\textbf{n}) \left[ \cos \left( \sum_{k=1}^d n_k \hat{i}_k \right) + \cos \left( \sum_{k=1}^d n_k \hat{j}_k \right) \right]\\
    \omega &= \sum_{\textbf{n} \in \mathcal{N}} \psi(\textbf{n}) \left[ \sin \left( \sum_{k=1}^d n_k \hat{i}_k \right) + \sin \left( \sum_{k=1}^d n_k \hat{j}_k \right) \right]\\
    \alpha_{\textbf{a}} &= \cos \left(\sum_{k=1}^d \left[ (i_k - a_k) \hat{i}_k + (j_k-a_k) \hat{j}_k \right]\right)\\
    \beta_{\textbf{a}} &= \sin \left( \sum_{k=1}^d \left[ (i_k-a_k)\hat{i}_k + (j_k-a_k) \hat{j}_k \right] \right)
  \end{aligned}
\end{equation}
Similarly, Eq. \eqref{eq:W} can be rewritten as,
\begin{equation}\label{eq:W_real}
  W_{\textbf{i},\textbf{j}} = -\frac{1}{(2\pi)^{2d}} \sum_{v_{\textbf{a}} \in \mathcal{D}} \int_{-\pi}^{\pi} \int_{-\pi}^{\pi} \frac{\alpha_{\textbf{a}} \sigma + \beta_{\textbf{a}} \omega}{\sigma^2 + \omega^2} d\hat{\textbf{i}} d\hat{\textbf{j}}
\end{equation}
The expressions in Eqs. \eqref{eq:Wt_real} and \eqref{eq:W_real} lend themselves to multi-dimensional numerical integration as the imaginary portion has been removed.
\section{Nearest Neighbor Lattice}
%
%
%
The one-dimensional ($d=1$), nearest neighbor ($\mathcal{N} = \{0,1,-1\}$) lattice (NNOD lattice) has the property that the shortest path between nodes $v_i$ and $v_j$ is $|i-j|$.
An example of this lattice is shown in Fig. \ref{fig:lattice}(A).
Let the edge weights be $\psi(1) = \psi(-1) = s > 0$ and $\psi(0) = p < -2s$.
The lattice function of the NNOD lattice is $\phi(\hat{i}) = p + 2s \cos \hat{i} < 0$.
As the Gramian entries are linear with respect to the contribution of each of the driver nodes, we examine each driver node separately so we assume $n_d = 1$.
We shift the indices $i' = i-a$ and $j'=j-a$, to place the driver node at $a' = 0$.
With this offset in mind, let $G_{i',j'} = W_{i,j}$ represent the shifted Gramian entries.
Equation \eqref{eq:W_real} for the NNOD lattice can be written as,
\begin{equation}\label{eq:Wnn}
  G_{i',j'} = \frac{-1}{\pi^2} \int_{0}^{\pi} \int_{0}^{\pi} \frac{\cos(i'\hat{i}) \cos(j'\hat{j})}{2p + 2s \cos \hat{i} + 2s \cos \hat{j}} d\hat{i} d\hat{j}.
\end{equation}
An analogous version of the double integral in Eq. \eqref{eq:Wnn} has been derived separately in the context of solving the discretized Helmholtz equation \cite{katsura1971lattice,morita1971useful}.\\
\indent
\begin{lem}[Recursion for Diagonal Values \cite{morita1971useful}]
  Let $\alpha = p^2/(2s^2)-1$, which, if the system is stable corresponds to a value of $\alpha > 1$.
  The diagonal values, $G_{i',i'}$, can be found from the recursion relation,
  \begin{equation}\label{eq:rr1}
    G_{i'+1,i'+1} = \frac{4i'}{2i'+1} \alpha G_{i',i'} - \frac{2i'-1}{2i'+1} G_{i'-1,i'-1}
  \end{equation}
  with initial values,
  \begin{equation}\label{eq:init}
    \begin{aligned}
      G_{0,0} &= \frac{1}{\pi |p|} K \left( \frac{4s^2}{p^2} \right)\\
      G_{1,1} &= \left( \frac{|p|}{2 \pi s^2} - \frac{1}{\pi |p|} \right) K \left( \frac{4s^2}{p^2} \right) - \frac{|p|}{2\pi s^2} E \left( \frac{4s^2}{p^2} \right) 
    \end{aligned}
  \end{equation}
  where $K(\cdot)$ and $E(\cdot)$ are the first and second complete elliptic integrals, respectively.
\end{lem}
\begin{cor}
  Let $z_{i'} = G_{i',i'} / G_{i'-1,i'-1}$ for $i' > 0$ represent the instantaneous rate of decay along the diagonal.
  The recursion in Eq. \eqref{eq:rr1} can be rewritten in terms of $z_{i'}$.
  \begin{equation}
    z_{i'+1} z_{i'} = \frac{4i'}{2i'+1} \alpha z_{i'} - \frac{2i'-1}{2i'+1}
  \end{equation}
  For large $i'$, this yields the approximate solution,
  \begin{equation}
    \tilde{z} = \alpha - \sqrt{\alpha^2-1},
  \end{equation}
  so that the asymptotic rate of decay of the diagonal elements is exponential,
  \begin{equation}\label{eq:scaling}
    G_{i',i'} \sim (\tilde{z})^{i'}.
  \end{equation}
\end{cor}
\begin{figure}
  \centering
  \includegraphics[width=\columnwidth]{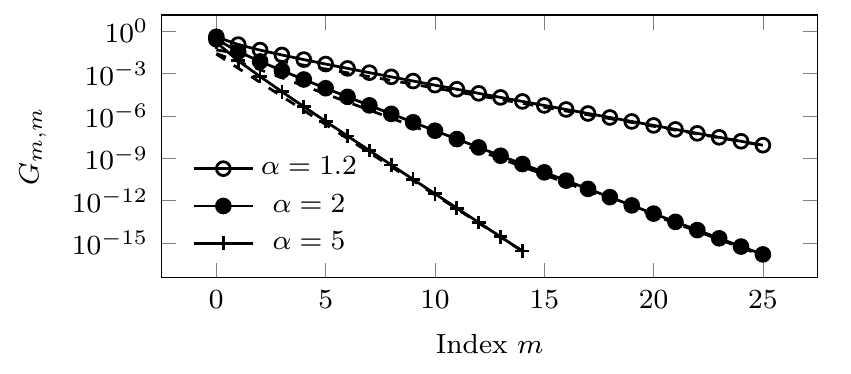}
  \caption{Asymptotic behavior of the diagonal values of the controllability Gramian of the NNOD lattice.
  The marks represent the values computed by numerically integrating Eq. \eqref{eq:Wnn} using two dimensional Gauss quadrature and the dashed lines represent the scaling in Eq. \eqref{eq:scaling}.
  We see that even for small values of $\alpha$ the asymptotic behavior provides a good approximation for small values of the diagonal index $m$.}
  \label{fig:diag}
\end{figure}
In Fig. \ref{fig:diag} we show that for both large and small values of $\alpha$, Eq. \eqref{eq:scaling} does a satisfactory job approximating $G_{i',i'}$ even for small values of $i'$.
%
\subsection{Control Metrics}
%
Two particularly useful control metrics are the trace of the inverse of the output Gramian and the logarithm of the determinant of the output Gramian \cite{summers2016submodularity}.
We assume the system is output controllable, so that $\bar{W}$ is positive definite and order the eigenvalues $0 < \lambda_1 \leq \lambda_2 \leq \ldots \leq \lambda_{n_t}$.
As we will see next, with knowledge of the diagonal elements, we can lower bound the control metrics using the Cauchy interlacing theorem \cite{golub2012matrix}.
%
\begin{thm}[Cauchy Interlacing Theorem \cite{golub2012matrix}]
  Let $X$ and $Y$ be two symmetric matrices of size $n$ and $m$ respectively, $n > m$, such that $Y$ is a principal submatrix of $X$.
  Order their eigenvalues such that $\lambda_i(X) \leq \lambda_{i+1}(X)$ and $\lambda_i(Y) \leq \lambda_{i+1}(Y)$.
  Then, each eigenvalue of $X$ can be bounded by,
  \begin{equation}
    \lambda_i(Y) \leq \lambda_i(X) \leq \lambda_{n-m+i}(Y), \quad i = 1,\ldots,m
  \end{equation}
\end{thm}
%
Also helpful is the Gerschgorin disc theorem \cite{golub2012matrix} as it provides a way to compute an upper bound to the largest eigenvalue of $\bar{W}$ for any given choice of the driver nodes.
\begin{thm}[Gerschgorin Disc Theorem \cite{golub2012matrix}]
  Let $A = \{A_{i,j}\}$ be an $n \times n$ square matrix.
  The eigenvalues of $A$, $\lambda(A)$, lie in the union of the $n$ discs $D_i$, $i = 1, \ldots, n$ in the complex plane, each centered at $A_{i,i}$ with radius $C_i = \sum_{j=1,j\neq i}^n A_{i,j}$.
  The largest eigenvalue of the matrix $A$ can thus be upper bounded by,
  \begin{equation}
    \lambda_{\max}(A) \leq \max\limits_{1<i<n} \sum_{j=1}^n A_{i,j}
  \end{equation}
\end{thm}
\begin{cor}
  It is not difficult to show from Eqs. \eqref{eq:Wnn} and \eqref{eq:init} that $G_{0,0} > G_{i',j'}$ for $|i'|+|j'| > 0$.
  Then, the largest eigenvalue of $\bar{W}$ for any choice of the driver node, $v_a$, can be upper bounded by,
  \begin{equation}
    \lambda_{\max}(\bar{W}) \leq \max\limits_{v_i \in \mathcal{T}} \sum_{v_j \in \mathcal{T}} G_{i',j'} \leq n_t G_{0,0}
  \end{equation}
  where we use the shift $i' = i-a$ and $j'=j-a$.
\end{cor}
\begin{figure}
  \centering
  \includegraphics[width=\columnwidth]{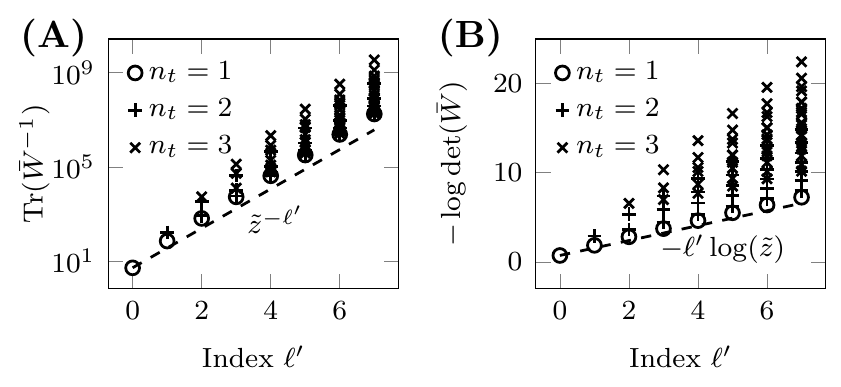}
  \caption{
    The lower bound of the control metrics as a function of index of node further from the driver node.
    The NNOD lattice has edge weights $p = 3$ and $s = 1$ and driver node $\mathcal{D} = \{v_0\}$.
    The set of target nodes $\mathcal{T} \subset \{v_0,v_1,\ldots,v_7\}$.
    (A) The trace of the inverse of the output controllability Gramian for all sets of target nodes with cardinality $|\mathcal{T}|=n_t = 1,2,3$.
    The dashed line represents Eq. \eqref{eq:trinv}.
    (B) the negative log determinant of the output controllability Gramian for all sets of target nodes with cardinality $|\mathcal{T}| = n_t = 1,2,3$.
    The dashed line represents Eq. \eqref{eq:logdet}.
  }
  \label{fig:scale}
\end{figure}
%
\subsubsection{Inverse of the Trace}
Let $\mathcal{B} = \left\{\left. \textbf{y} \in \mathcal{R}^{n_t} \right| ||\textbf{y}|| = 1 \right\}$ be the $n_t$-dimensional ball centered at the origin.
The average minimum energy required to reach a point $\textbf{b}$, located on the unit hypersphere, from the origin, is found by integrating over the $\mathcal{B}$ \cite{summers2014optimal,summers2016submodularity}.
\begin{equation}
  \frac{\int_{\mathcal{B}} \textbf{y}^T \bar{W}^{-1} \textbf{y} d\textbf{y}}{\int_{\mathcal{B}} d\textbf{y}} = \frac{1}{n_t} \text{Tr} (\bar{W}^{-1})
\end{equation}

%
%
%
Let $n_t > 1$ and $v_{\ell'} \in \mathcal{T}$ such that $\ell' = \max\limits_{v_{i'} \in \mathcal{T}} i'$, i.e., the index of the node furthest from the single driver node.
Using the Cauchy interlacing theorem which states that $\lambda_1(\bar{W}) < G_{\ell',\ell'}$, we can lower bound the trace of the inverse of the output controllability Gramian.
\begin{equation}\label{eq:trinv}
  \begin{aligned}
    \text{Tr}(\bar{W}^{-1}) &\geq \frac{1}{\lambda_1(\bar{W})} \geq \frac{1}{G_{\ell',\ell'}} \sim \tilde{z}^{-\ell'}
  \end{aligned}
\end{equation}
From the above expression we see that, as $0 < \tilde{z} < 1$, by reducing $\ell'$, we can reduce the lower bound.
An example of the lower bound in Eq. \eqref{eq:trinv} is shown in Fig. \ref{fig:scale}(A) for a NNOD lattice with edge weights $p = 3$ and $s = 1$.
Namely, the figure shows that by reducing $\ell'$ one can exponentially decrease the trace of the inverse output Controllability Gramian.
This indicates that a convenient choice of an alternate driver node is the one that reduces the distance to the furthest target node.
\subsubsection{Volume of the Controllability Ellipsoid}
The controllability ellipsoid is defined as the set of vectors $\textbf{y}$ such that,
\begin{equation}
  \mathcal{S} = \left\{ \textbf{y} \in \mathcal{R}^{n_t} | \textbf{y}^T \bar{W}^{-1} \textbf{y} \leq 1\right\}
\end{equation}
The volume of $\mathcal{S}$ \cite{summers2016submodularity} is
\begin{equation}\label{eq:volume}
  V(\bar{W}) = \frac{\pi^{n_t/2}}{\Gamma(n_t/2+1)} \left(\det \bar{W}\right)^{1/n_t}
\end{equation}
Note that $cV(\bar{W}) = V(c\bar{W})$, where, from Gerschgorin theorem we see that by setting $c = \frac{1}{n_t G_{0,0}}$, $\lambda_{n_t} (c\bar{W}) < 1$.
Minimizing $-\log \det (c\bar{W})$ can be interpreted as maximizing the volume of the reachable subspace for any value $J^*$.
\begin{equation}\label{eq:logdet}
  \begin{aligned}
    -\log\det (c\bar{W}) &\geq -\log (\lambda_1 (c\bar{W}))\\
    &\geq - \log (cG_{\ell',\ell'}) \sim - \log (\tilde{z}) \ell'
  \end{aligned}
\end{equation}
As $0 < \tilde{z} < 1$, $-\log(\tilde{z}) > 0$, and so the lower bound of -$\log \deg c (\bar{W})$ scales with $\ell'$.
An example of the lower bound in Eq. \eqref{eq:logdet} is shown in Fig. \ref{fig:scale}(B).
Namely, the figure shows that the bound holds tightly for $n_t = 1$ but becomes more conservative as the number of targets increase.
By selecting a driver node such that $\ell'$ is minimized the $\log \det (\bar{W})$ may exponentially decrease the minimum control energy.
  %
  %
  %
  %
%
\section{Discussion and Conclusion}
We have derived the exact expressions of the controllability Gramian for networked systems where the underlying topology is a lattice.
We have specialized the results to NNOD lattices so that the length of the shortest path between two nodes appears in the expression for the Gramian entries.
Our analytical expressions are in agreement with previously reported observations \cite{chen2016energy,wang2017physical}, that control energy metrics scale exponentially with respect to distance metrics for finite graphs with general topology.
Current research into driver node selection methods 
require the computation of many controllability Gramians (and their log determinant or inverse trace) which becomes prohibitively expensive for large networks in terms of both computation and storage.
Instead, the positive correlation between distance metrics with respect to the driver nodes and target nodes and the energy metrics discussed suggests that one may construct a heuristic driver node selection method based purely on the topology of the underlying graph.
%
An exhaustive numerical comparison between the greedy approximation algorithms \cite{summers2016submodularity,tzoumas2016minimal} and heuristic algorithms that minimize a distance metric between the sets of driver nodes and target nodes is forthcoming in a future publication.
\begin{ack}                               
  This work is supported by the National Science Foundation through NSF grant CMMI-1400193, NSF grant CRISP-1541148, and ONR Award No. N00014-16-1-2637 as well as HDTRA1-13-1-0020.
\end{ack}
\bibliographystyle{plain}

\begin{thebibliography}{10}

\bibitem{benner2013numerical}
Peter Benner and Jens Saak.
\newblock Numerical solution of large and sparse continuous time algebraic
  matrix riccati and lyapunov equations: a state of the art survey.
\newblock {\em GAMM-Mitteilungen}, 36(1):32--52, 2013.

\bibitem{chen2016energy}
Yu-Zhong Chen, Le-Zhi Wang, Wen-Xu Wang, and Ying-Cheng Lai.
\newblock Energy scaling and reduction in controlling complex networks.
\newblock {\em Royal Society open science}, 3(4):160064, 2016.

\bibitem{cowan2012nodal}
Noah~J. Cowan, Erick~J. Chastain, Daril~A. Vilhena, James~S. Freudenberg, and
  Carl~T. Bergstrom.
\newblock Nodal dynamics, not degree distributions, determine the structural
  controllability of complex networks.
\newblock {\em PloS one}, 7(6):e38398, 2012.

\bibitem{gao2014target}
Jianxi Gao, Yang-Yu Liu, Raissa~M. D'souza, and Albert-L{\'a}szl{\'o}
  Barab{\'a}si.
\newblock Target control of complex networks.
\newblock {\em Nature communications}, 5:5415, 2014.

\bibitem{golub2012matrix}
Gene~H. Golub and Charles~F. Van~Loan.
\newblock {\em Matrix computations}, volume~3.
\newblock JHU Press, 2012.

\bibitem{katsura1971lattice}
Shigetoshi Katsura, Tohru Morita, Sakari Inawashiro, Tsuyoshi Horiguchi, and
  Yoshihiko Abe.
\newblock Lattice green's function. introduction.
\newblock {\em Journal of Mathematical Physics}, 12(5):892--895, 1971.

\bibitem{klickstein2017energy}
Isaac Klickstein, Afroza Shirin, and Francesco Sorrentino.
\newblock Energy scaling of targeted optimal control of complex networks.
\newblock {\em Nature communications}, 8:15145, 2017.

\bibitem{lin1974structural}
Ching-Tai Lin.
\newblock Structural controllability.
\newblock {\em IEEE Transactions on Automatic Control}, 19(3):201--208, 1974.

\bibitem{lin2011augmented}
Fu~Lin, Makan Fardad, and Mihailo~R. Jovanovic.
\newblock Augmented lagrangian approach to design of structured optimal state
  feedback gains.
\newblock {\em IEEE Transactions on Automatic Control}, 56(12):2923--2929,
  2011.

\bibitem{liu2016control}
Yang-Yu Liu and Albert-L{\'a}szl{\'o} Barab{\'a}si.
\newblock Control principles of complex systems.
\newblock {\em Reviews of Modern Physics}, 88(3):035006, 2016.

\bibitem{liu2011controllability}
Yang-Yu Liu, Jean-Jacques Slotine, and Albert-L{\'a}szl{\'o} Barab{\'a}si.
\newblock Controllability of complex networks.
\newblock {\em Nature}, 473(7346):167, 2011.

\bibitem{mirchandani1990discrete}
Pitu~B. Mirchandani and Richard~L. Francis.
\newblock {\em Discrete location theory}.
\newblock John Wiley \& Sons, Inc, 1990.

\bibitem{morita1971useful}
Tohru Morita.
\newblock Useful procedure for computing the lattice green's function-square,
  tetragonal, and bcc lattices.
\newblock {\em Journal of mathematical physics}, 12(8):1744--1747, 1971.

\bibitem{olshevsky2014minimal}
Alex Olshevsky.
\newblock Minimal controllability problems.
\newblock {\em IEEE Transactions on Control of Network Systems}, 1(3):249--258,
  2014.

\bibitem{pequito2017robust}
S{\'e}rgio Pequito, Guilherme Ramos, Soummya Kar, A.~Pedro Aguiar, and Jaime
  Ramos.
\newblock The robust minimal controllability problem.
\newblock {\em Automatica}, 82:261--268, 2017.

\bibitem{summers2016actuator}
Tyler Summers.
\newblock Actuator placement in networks using optimal control performance
  metrics.
\newblock In {\em Decision and Control (CDC), 2016 IEEE 55th Conference on},
  pages 2703--2708. IEEE, 2016.

\bibitem{summers2016submodularity}
Tyler~H. Summers, Fabrizio~L. Cortesi, and John Lygeros.
\newblock On submodularity and controllability in complex dynamical networks.
\newblock {\em IEEE Transactions on Control of Network Systems}, 3(1):91--101,
  2016.

\bibitem{summers2014optimal}
Tyler~H. Summers and John Lygeros.
\newblock Optimal sensor and actuator placement in complex dynamical networks.
\newblock {\em IFAC Proceedings Volumes}, 47(3):3784--3789, 2014.

\bibitem{sun2013controllability}
Jie Sun and Adilson~E. Motter.
\newblock Controllability transition and nonlocality in network control.
\newblock {\em Physical review letters}, 110(20):208701, 2013.

\bibitem{tang2014synchronization}
Yang Tang, Feng Qian, Huijun Gao, and J{\"u}rgen Kurths.
\newblock Synchronization in complex networks and its application--a survey of
  recent advances and challenges.
\newblock {\em Annual Reviews in Control}, 38(2):184--198, 2014.

\bibitem{tzoumas2015minimal}
Vasileios Tzoumas, Mohammad~Amin Rahimian, George~J. Pappas, and Ali Jadbabaie.
\newblock Minimal actuator placement with optimal control constraints.
\newblock In {\em American Control Conference (ACC), 2015}, pages 2081--2086.
  IEEE, 2015.

\bibitem{tzoumas2016minimal}
Vasileios Tzoumas, Mohammad~Amin Rahimian, George~J. Pappas, and Ali Jadbabaie.
\newblock Minimal actuator placement with bounds on control effort.
\newblock {\em IEEE Transactions on Control of Network Systems}, 3(1):67--78,
  2016.

\bibitem{wang2015passivity}
Jin-Liang Wang, Huai-Ning Wu, and Tingwen Huang.
\newblock Passivity-based synchronization of a class of complex dynamical
  networks with time-varying delay.
\newblock {\em Automatica}, 56:105--112, 2015.

\bibitem{wang2017physical}
Le-Zhi Wang, Yu-Zhong Chen, Wen-Xu Wang, and Ying-Cheng Lai.
\newblock Physical controllability of complex networks.
\newblock {\em Scientific reports}, 7:40198, 2017.

\bibitem{yan2012controlling}
Gang Yan, Jie Ren, Ying-Cheng Lai, Choy-Heng Lai, and Baowen Li.
\newblock Controlling complex networks: How much energy is needed?
\newblock {\em Physical review letters}, 108(21):218703, 2012.

\bibitem{yan2015spectrum}
Gang Yan, Georgios Tsekenis, Baruch Barzel, Jean-Jacques Slotine, Yang-Yu Liu,
  and Albert-L{\'a}szl{\'o} Barab{\'a}si.
\newblock Spectrum of controlling and observing complex networks.
\newblock {\em Nature Physics}, 11(9):779, 2015.

\bibitem{yu2010some}
Wenwu Yu, Guanrong Chen, and Ming Cao.
\newblock Some necessary and sufficient conditions for second-order consensus
  in multi-agent dynamical systems.
\newblock {\em Automatica}, 46(6):1089--1095, 2010.

\bibitem{yu2009pinning}
Wenwu Yu, Guanrong Chen, and Jinhu L{\"u}.
\newblock On pinning synchronization of complex dynamical networks.
\newblock {\em Automatica}, 45(2):429--435, 2009.

\bibitem{yuan2013exact}
Zhengzhong Yuan, Chen Zhao, Zengru Di, Wen-Xu Wang, and Ying-Cheng Lai.
\newblock Exact controllability of complex networks.
\newblock {\em Nature communications}, 4:2447, 2013.

\bibitem{zhang2017efficient}
Xizhe Zhang, Huaizhen Wang, and Tianyang Lv.
\newblock Efficient target control of complex networks based on preferential
  matching.
\newblock {\em PloS one}, 12(4):e0175375, 2017.

\bibitem{zhao2005onset}
Liang Zhao, Ying-Cheng Lai, Kwangho Park, and Nong Ye.
\newblock Onset of traffic congestion in complex networks.
\newblock {\em Physical Review E}, 71(2):026125, 2005.

\bibitem{zhou2015controllability}
Tong Zhou.
\newblock On the controllability and observability of networked dynamic
  systems.
\newblock {\em Automatica}, 52:63--75, 2015.

\end{thebibliography}

\end{document}